\documentstyle[12pt]{article}
\def\bg{\bar{\Gamma}}
\def\bpsi{\bar{\psi}}
\def\dfrac{\displaystyle\frac}
\def\i{\imath}

\def\de{\delta}
\def\a{\alpha}
\def\hp{\hat{p}}
\def\hk{\hat{k}}

\def\bp{{\bf p}}

\def\d{\partial}

\def\noi{\noindent}
\def\ve{\varepsilon}
\newcommand{\Eq}[1]{Eq.~(\ref{#1})}

\newcommand{\refc}[1]{Ref.~\cite{#1}}

\newcommand{\bea}{\begin{eqnarray}}
\newcommand{\eea}{\end{eqnarray}}
\newcommand{\be}{\begin{equation}}
\newcommand{\ee}{\end{equation}}
\newcommand{\bc}{\begin{center}}
\newcommand{\ec}{\end{center}}
\newcommand{\ba}{\begin{array}}
\newcommand{\ea}{\end{array}}
\newcommand{\non}{\nonumber}
\newcommand{\cL}{{\cal L}}

\newcommand{\annp}[3]{{\it  Ann. Phys. (N.Y.) }{{\bf #1} {(#2)} {#3}}}
\newcommand{\ijmp}[3]{{\it Int. J. Mod. Phys. } {{\bf #1} {(#2)} {#3}}}

\newcommand{\prd}[3]{{\it  Phys. Rev. D} {{\bf #1} {(#2)} {#3}}}

\newcommand{\sovjnp}[3]{{\it Sov. J. Nucl. Phys. }{{\bf #1} {(#2)} {#3}}}

\normalsize
\begin{document}

\large
\thispagestyle{empty}
\begin{flushright}
                                           FIAN/TD/97-28\\
                                                hep-th/9807015

\end{flushright}

{}
\bc
\vspace{2cm}

\normalsize
\phantom{q}

\vspace{2cm}

{\LARGE\bf Curing fermion mass gauge variance in QED$_{2+1}$}

\vspace{5ex}

{\Large I.~V.~Tyutin$^{\dagger}$ and Vadim Zeitlin$^{\ddagger}$}

{\large Department of Theoretical Physics, P.~N.~Lebedev Physical
Institute,

  Leninsky prospect 53, 117924 Moscow, Russia}
\vspace{5ex}

\ec

\vspace{1cm}
\centerline{{\Large\bf Abstract}}

\normalsize
\begin{quote}
The problem of the gauge dependence of the fermion mass in
the Maxwell-Chern-Simons QED$_{2+1}$  is revisited. Using
Proca mass term as an intermediate infrared regulator we are demonstrating
gauge-invariance of the fermion mass shell in QED$_{2+1}$ in all
orders of the perturbation theory.
\end{quote}

\vfill
\noindent
$^\dagger$ E-mail address: tyutin@td.lpi.ac.ru

\noindent
$^\ddagger$ E-mail address: zeitlin@td.lpi.ac.ru

%%%%%%%%%%%%%%%%%%%%%%%%%%%%%%%%%%%%%%%%%%%%%%%%%%%%%%%%%%%%%%%

\newpage
\setcounter{page}{1}

\section{Introduction}

Quantum field theories in 3 space-time dimensions provide fertile
ground for studying high-temperature asymptotics of 4-dimensional models
\cite{highT} and different effects in planar systems in condensed matter
\cite{condmat}. Comparing with
ordinary 4-dimensional theories the lower dimension looks as an apparent
simplification.  However, lower space-time dimension brings
infrared singularities that have grave consequences.  In one of the first
papers on 3-dimensional models \cite{DJT} it was indicated that such a
singularity leads to the gauge dependence of the one-loop fermion mass
shell. On the one-loop level a simple solution of the problem was proposed
 \cite{DJT}:
only the transversal (Landau) gauge $\a=0$ is valid, since it provides the
same result as the Coulomb gauge (which is actually ill-defined),
but in higher orders of the perturbation theory the situation is
uncertain. Moreover, gauge variance of the fermion mass shell may indicate
presence of an anomaly even in the transversal gauge.

In this
letter we shall discuss application of the Proca regularization to the
topologically massive QED$_{2+1}$. We shall demonstrate that in the
Proca-Maxwell-Chern-Simons QED$_{2+1}$  (Proca model) in a
relativistic gauge the Ward identities imply gauge-invariance of the fermion
mass shell. The regularization may be taken off and in the limit of zero
Proca mass the model reduces to the topologically massive QED$_{2+1}$ in the
transversal gauge. We shall demonstrate the infrared finiteness of the
topologically massive QED$_{2+1}$ in the transversal gauge in
all orders of the perturbation theory and prove the unitarity of the model.

\bigskip
The Lagrangian of the Proca model is the following:

        \be
        \cL = -\frac14 F_{\mu\nu}F^{\mu\nu} +
        \frac{\theta}4 \ve_{\mu\nu\alpha}F^{\mu\nu}A^\alpha
        + \frac{m^2}2 A_\mu A^\mu- \frac1{2\alpha} (\d_\mu A^\mu)^2+
        \bar{\psi}
        (\imath {\partial  \kern-0.5em/} + e {A\kern-0.5em/}   -M)\psi~~~.
        \label{mcs}
        \ee
\noi
We are using standard notations, $\theta$ is the Chern-Simons mass and $m$
is the Proca mass.  The corresponding gauge field propagator is

	 \be
        D_{\mu\nu}^{\rm Proca}(p) =
        -\i\frac{p^2 -m^2}{(p^2 - m^2)^2-\theta^2 p^2}
        \left(
                \left( g_{\mu\nu} - \frac{p_\mu p_\nu}{p^2} \right)
                + \i \theta \ve_{\mu\nu\alpha}\dfrac{p^\alpha}{p^2-m^2}
							  \right)
               -\i\frac{p_\mu p_\nu}{p^2}
               \frac{\alpha}{p^2-\alpha m^2}
        \label{dmunu}
        \ee

\noi
(at $m \to 0$ the latter is a propagator of the topologically massive
QED$_{2+1}$ in a relativistic gauge $\a$).

The Ward identities in the Proca model have the standard form:

        \be
        \d_\mu \dfrac{\delta\bar{\Gamma}}{\delta A_\mu(x)}=
        \i e \psi(x) \dfrac{\delta\bar{\Gamma}}{\delta \psi(x)}
        -\i e \bar{\psi}(x) \dfrac{\delta\bar{\Gamma}}{\delta
        \bar{\psi}(x)}~~~,
        \label{ward1}
        \ee

        \be
        \Gamma =
        \bar{\Gamma} - \frac1{2\a} (\d_\mu A^\mu)
        (\d_\nu A^\nu) +\frac{m^2}2A_\mu A^\mu~~~
        =  W - IA - \eta\psi - \bar{\eta}\bar{\psi},
        \non
        \ee

        \be
        e^{\i W(I,\eta,\bar{\eta})}=\int D\!A D\!\psi D\!\bar{\psi}
        e^{\i \int d\!x ({\cal L} +IA +\eta\psi + \bar{\eta}\bar{\psi})}~~~.
        \ee

Taking \Eq{ward1} into account one may obtain the derivative of the generating
functional of the vertex functions $\bg$,

        \be
        \frac{\delta \bar{\Gamma}}{\delta \alpha} =
        \frac{e}{2\alpha}
        \int d\!x d\!y
        \left[
        \d_\mu \frac{\de^2W}{\de I^\mu(x)\de {\eta}(y)}
        {\rm D} (x-y) \frac{\delta \bar{\Gamma}}{\delta {\psi}(y)}
        -
        \d_\mu \frac{\de^2W}{\de I^\mu(x) \de \bar{\eta}(y)}
        {\rm D} (x-y) \frac{\delta \bar{\Gamma}}{\delta \bar{\psi}(y)}
                                                \right]~~,
        \label{varG}
        \ee

        \be
        {\rm D}(x) = \int \frac{d\! p}{(2\pi)^3}
        \frac{e^{-\i px}}{p^2-\a m^2}~~~.
        \label{romanD}
        \ee

Calculating variations of ${\delta \bar{\Gamma}}/{\delta
\alpha}$ with respect to $A, \psi, \bpsi$ one may obtain derivatives of the
corresponding vertex functions with respect to the gauge parameter $\a$. For
instance, the derivative of the fermion mass operator $\Sigma$ is the
following:

        \bea
        \lefteqn{\frac{\d \Sigma (x-y)}{\d \a} =
        -\i \frac{e}2 \int d\!z
        \left[
        \int d\!z_1 d\!z_2 D^\mu (z-z_1) \Gamma_\mu(z_1,x,z_2)
        S(z_2-z)                \right]
        S^{-1} (z-y) -}~~~~~~~~~~~~\nonumber\\
        &&~~~~~-\i \frac{e}2 \int d\!z
        S^{-1} (x-z)
        \left[
        \int d\!z_1 d\!z_2 D^\mu (z-z_2) S(z-z_1)
        \Gamma_\mu(z_1,z_2,y) \right]~~~,
        \label{varSigma}
        \eea

        \be
        \Gamma^\mu(x,y,z) =
        \left.
        \frac{\de}{\de A_\mu(x)}
        \frac{\de}{\de \bar{\psi(y)}}
        \frac{\de}{\de {\psi(z)}} \Gamma
        \right|_{€_\mu=\psi=\bar{\psi}=0}~~~,
        \ee

\noi
$S$ is an exact fermion propagator, the mass operator  ~$\Sigma$~ is defined
as ~$S^{-1} = S^{-1}_0 +\i \Sigma$, ~$S_0$ is the bare fermion propagator,
and the function $D_\mu$ is the following:

        \be
        D_\mu(x) = \int \frac{d\!p}{(2\pi)^3} e^{-\i px}
        \frac{p_\mu}{(p^2-\a m^2)^2}~~~.
        \ee

In the one-loop approximation

        \bea
        \lefteqn{\frac{\d \Sigma_1 (x-y)}{\d \a} =}\\
        \nonumber
        &&-\i e^2 \int d\!zS^{-1}_0(x-z) [D_\mu(x-y)S_0(x-y)\gamma^\mu]
        \equiv
        \i e^2 \int d\!zS^{-1}_0(x-z) \Sigma'_1(z-y)~~~.
        \eea

In the momentum space the latter may be rewritten as follows:

        \be
        \frac{\d \Sigma_1 (p)}{\d \a} = e^2 (\hp - M) {\Sigma}'_1(p)~~~,
        \label{sigma'}
        \ee

        \be
        {\Sigma}'_1 (p) = -\frac{\i}{(2\pi)^3} \int d\!k
        \frac{(\hat{p}-\hat{k}+M)\hat{k}}{(k^2-\alpha
        m^2)^2((p-k)^2-M^2)}~~~.
        \ee

\smallskip
On the fermion mass shell the quantity $\Sigma'_1$ is nonsingular,
${\Sigma}'_1 (\hp=M) = \frac1{8\pi m \sqrt{\a}}$ and \Eq{sigma'} implies
independence of the one-loop fermion mass of the gauge parameter in the Proca
model\footnote{In the topologically massive QED$_{2+1}$ the corresponding
expression for ${\Sigma}'_1$ is ${\Sigma}'_1 (p) = \frac1{8\pi}\cdot
\frac1{\hp -M} + \dots$ , where $\dots$ denotes terms logarithmically
divergent and finite on the mass shell, and the gauge invariance of the
one-loop fermion mass does not follow from the Ward identities.}. Therefore,
the one-loop fermion mass renormalization does not depend on the gauge
parameter and in the $m \to 0$ limit it coincides with the mass
renormalization in the topologically massive QED$_{2+1}$ in $\a=0$ gauge. We
shall show below that in the Proca model all the vertex functions on the
fermion mass shell are independent of the gauge parameter, and transversal in
the photon momenta.

Transversality of vertex functions with respect to  external photon momenta
follows from \Eq{ward1} providing that the corresponding
diagram is nonsingular on the fermion mass shell. The gauge invariance of the
vertex functions follows from \Eq{varG}.
To prove the gauge invariance of the model, which is equivalent to
establishing the fact that \Eq{varG} vanishes on the mass shell,
we shall use the approach elaborated in \refc{KT}.  The right-hand side of
\Eq{varG} has the following structure:

\medskip
\begin{equation}
\int \!dp \psi_{in}(p)\vec{K}(p)\int\! dk
\begin{picture}(100.00,40.00)
\unitlength=1.00mm
\linethickness{0.4pt}
\put(-41,-14){
\put(44.00,15.){\line(1,0){20.}}
\put(59.00,15.){\vector(1,0){1.}}
\put(70.00,20.00){\oval(15.,15.)}
\put(74.00,27.){\line(1,1){7.}}
\put(78.00,20.){\line(1,0){10.}}
\put(74.00,13.){\line(1,-1){7.}}
\put(44.,15.){\circle*{1.}}
\put(70.00,20.){\makebox(0,0)[cc]{{$M$}}}
\put(55.00,12.){\makebox(0,0)[cc]{$p-k$}}
\put(46.,16.5){\circle*{0.5}}
\put(48.,18.2){\circle*{0.5}}
\put(50.,19.3){\circle*{0.5}}
\put(52.,20.4){\vector(2,1){0.5}}
\put(52.,20.4){\circle*{0.5}}
\put(54.,21.4){\circle*{0.5}}
\put(56.,22.3){\circle*{0.5}}
\put(58.,23.1){\circle*{0.5}}
\put(60.,23.9){\circle*{0.5}}
\put(62.,24.5){\circle*{0.5}}
\put(53.00,25.){\makebox(0,0)[cc]{$k$}}
}
\end{picture}~~~~~~~~~~~~~~~~~~~,~~~~~~~~~~~~~~
\label{fig1}
\end{equation}

\bigskip
\noi
where the dotted line denotes the propagator $\frac{k_\mu}{k^2(k^2-\theta^2)}$
, ~$K(p)= \hp -M$ , ~$\psi_{in}(p)$ is a solution of the free equation of
motion, $K(p)\psi_{in}(p)= 0$ , $\psi_{in}(p) \sim \de (p^2- M^2)$.

\Eq{fig1} vanishes if $K(p)$ acts on a function that has no $\frac1{\hp -
M}$ singularities. If the diagram in the right-hand side of \Eq{fig1} is
one-particle irreducible, the latter equation may be rewritten as

        \be
        \int \! dk \frac{(\hp-\hk+M))k_\mu}
        {(k^2-2pk+\de)k^2(k^2-\theta^2)}M_\mu(k)~~~,~~~ \de=p^2-M^2~~~~,
        \label{irred}
        \ee

\smallskip
\noi
and it vanishes indeed on the fermion mass shell.

If the diagram \Eq{fig1} is a reducible one,

\bigskip
\begin{equation}
\begin{picture}(100.00,40.00)
\unitlength=1.00mm
\linethickness{0.4pt}
\put(-70,-18){
\put(45.00,20.){\line(1,0){17.}}
\put(54.00,20.){\vector(1,0){1.}}
\put(70.00,20.00){\oval(15.,15.)}
\put(74.00,27.){\line(1,1){7.}}
\put(77.00,20.){\line(1,0){10.}}
\put(74.00,13.){\line(1,-1){7.}}
\put(70.00,20.){\makebox(0,0)[cc]{{$M'$}}}
\put(40.00,20.){\makebox(0,0)[cc]{{$\gamma$}}}
\put(53.00,15.){\makebox(0,0)[cc]{$p$}}
\put(45,20){\circle*{1.}}
\put(35,20){\circle*{1.}}
\put(40.00,20.00){\oval(10.,10.)}
}
\end{picture}
\begin{picture}(100.00,40.00)
\unitlength=1.00mm
\linethickness{0.4pt}
\put(-41,-14){
\put(44.00,15.){\line(1,0){20.}}
\put(59.00,15.){\vector(1,0){1.}}
\put(70.00,20.00){\oval(15.,15.)}
\put(76.00,15.){\line(1,0){10.}}
\put(44.,15.){\circle*{1.}}
\put(70.00,20.){\makebox(0,0)[cc]{{$\Gamma$}}}
\put(35.00,17.){\makebox(0,0)[cc]{,~~$\gamma=$}}
\put(55.00,12.){\makebox(0,0)[cc]{$p-k$}}
\put(46.,16.5){\circle*{0.5}}
\put(48.,18.2){\circle*{0.5}}
\put(50.,19.3){\circle*{0.5}}
\put(52.,20.4){\vector(2,1){0.5}}
\put(52.,20.4){\circle*{0.5}}
\put(54.,21.4){\circle*{0.5}}
\put(56.,22.3){\circle*{0.5}}
\put(58.,23.1){\circle*{0.5}}
\put(60.,23.9){\circle*{0.5}}
\put(62.,24.5){\circle*{0.5}}
\put(53.00,25.){\makebox(0,0)[cc]{$k$}}
}
\end{picture}
\label{fig2}
\end{equation}

\vspace{1.1cm}
\noi
the equation similar to \Eq{irred} is nonzero. However, its contribution will
be cancelled identically after the fermion wave function renormalization
\cite{KT} (note that in the topologically massive QED$_{2+1}$ in an arbitrary
$\a$-gauge the dotted line in Eqs. (\ref{fig1}),(\ref{irred}) would be
${k_\mu}/{k^4}$, that makes \Eq{fig1} singular, even if the vertex function
exists). Transversality of the vertex functions may be proven along the same
lines.

Thus we have shown that the fermion mass shell in the Proca model is
gauge-invariant. Now, if we could take the regularization off,
we would obtain a limit corresponding to the
topologically massive QED$_{2+1}$ in $\a=0$ gauge, which would imply the
unitarity of the {\it regularized} topologically massive QED$_{2+1}$.

Let us show that $m \to 0$ limit of the Proca model is regular. The gauge
field in the Proca model has the following eigenvectors decomposition:

        \be
        A^{{\rm Proca}}_\mu(x)=
        \sum_{i=1}^{3}
        \int\frac{d {\bf p}}{2\pi\sqrt{2\omega_i}}
        (e^{-\i p^{(i)} x}u^{(i)}_\mu a_{i}(\bp) + {\rm h.c.})~~~,~~~
        p_\mu^{(i)}=(\omega_i,\bp)~~~,
        \label{pol_v1}
        \ee

        \be
        u^{(1,2)}_\mu =
        \dfrac{\omega_{1,2}}
        {|\bp| \sqrt{\omega_{1,2}^2-\bp^2+m^2}}
        (p^{(1,2)}_\mu - g_{\mu 0} \frac{\omega_{1,2}^2-\bp^2}{\omega_{1,2}}
        +\i \frac{\omega_{1,2}^2-\bp^2-m^2}{\theta\omega_{1,2}}
        \varepsilon_{\mu\alpha0}p^{(1,2)\alpha})~~~,
        \label{pol_v2}
        \ee

        \be
        \omega_{1,2}=\sqrt{{\bf p}^2+m^2_{1,2}}~~~,~~~
        m_{1,2}^2=
        m^2+\frac{\theta^2}2 \pm \theta\sqrt{m^2+\frac{\theta^2}4}~~~,
        \ee

        \be
        u^{(3)}_\mu= \frac1{m}p_\mu^{(3)} ~~~,~~~\omega_3=\sqrt{{\bf
        p}^2+m^2_3}~~~,~~~m_3^2=\alpha m^2~~~.
        \ee

Excitations $u^{(1)}_\mu$ and  $u^{(2)}_\mu$ are the physical modes, and
the excitation  $u^{(3)}_\mu$ is a nonphysical longitudinal mode.  Due to the
transversality of the vertex functions all the scattering amplitudes wherein
photons with the polarization $u^{(3)}_\mu$ are involved do not contribute to
the scattering matrix.

On the other hand, in the topologically massive QED$_{2+1}$ just one
excitation  is physical, namely $e^{(1)}_\mu$:

        \be
        A^{{\rm QED}}_\mu(x)=
        \int\frac{d {\bf p}}{2\pi\sqrt{2\omega\phantom{|}}}
        (e^{-\i p^{(1)}x}e^{(1)}_\mu a_{ph}(\bp) + {\rm h.c.})
        +
        \int\frac{d {\bf p}}{2\pi\sqrt{2|{\bf p}|}}
        (e^{-\i p^{(2)}x}(e^{(2)}_\mu a_2(\bp) + e^{(3)}_\mu a_3(\bp)) + {\rm
        h.c.})~~,
        \ee

        \be
        e^{(1)}_\mu = \frac{\omega}{\theta|\bp|}
        (p^{(1)}_\mu - g_{\mu 0} \frac{\theta^2}{\omega}
        +\i \frac\theta{\omega}\varepsilon_{\mu\alpha0}p^{(1)\alpha})
        ~~~~, ~~~ \omega = p_0^{(1)}=\sqrt{\theta^2+{\bf p}^2}~~~.
        \ee

        \be
        e^{(2)}_\mu=\frac{1}{\sqrt{\theta|\bp|}} p_\mu^{(2)},~~~
        p_0^{(2)}=p_0^{(3)}=|{\bf p}|
        \ee

        \be
        e^{(3)}_\mu = -\frac{1}{2}\sqrt{\frac{\theta}{|\bp|}}
        \left(\alpha g_{\mu 0} + \frac{2\i}{\theta}
        \varepsilon_{\mu\beta0}p^{(3)\beta}
        + \bigg(\frac{|\bp|}{\theta^2} - \frac{\alpha}{2|\bp|}
        - \i \alpha x_0 \bigg)   p^{(3)}_\mu
                                        \right)~~~,
        \label{pol_e3}
        \ee

\noi
$e^{(2)}_\mu$  and $e^{(3)}_\mu$ are massless nonphysical excitations.

In $m \to 0$ limit $u^{(1)}_\mu$ turns into $e^{(1)}_\mu$,
 and $u^{(2)}_\mu$ may be rewritten as follows:

        \be
        u^{(2)}_\mu|_{m\to 0} =
        \frac1{m}
        p_\mu + m l_\mu(p)~~~,
        \label{u2_m=0}
        \ee

\noi
the vector $l_\mu$ having a finite limit at $m \to 0$.
After taking the regularization off the contribution of the photons of
polarization $u^{(2)}_\mu$ to the scattering amplitudes should
vanish.  Due to the transversality
of vertices the contribution of the longitudinal part of $u^{(2)}_\mu$
vanishes and the remainder is proportional to $m$. Therefore, emitting  of
$n$ "soft" photons (of the polarization $u^{(2)}_\mu$) gives rise to the
coefficient  $m^n$.  Thus, if diagrams with ~$n$ ~$u^{(2)}_\mu$ photons
have infrared singularities lower than $m^n$, the corresponding scattering
matrix elements vanish in $m \to 0$ limit.

Fortunately, in the transversal gauge all diagrams are {\it infrared
finite}. The proof is  straightforward. In $\a=0$ gauge the infrared-dominant
term of the gauge field propagator is proportional to
$\ve_{\mu\nu\a} k^\a/k^2$. Consider a closed fermion loop with $n$ outgoing
photons, $\Pi_{\mu_1\dots\mu_n}(p_1,\dots,p_n)$. Due to the transversality,
$p^{\mu_i}\Pi_{\mu_1\dots\mu_i\dots\mu_n}(p_1,\dots,p_n)=0$, the
diagram $\Pi_{\mu_1\dots\mu_n}(p_1,\dots,p_n)$ for $n>2$ is proportional to
a momentum of {\it each} photon, thus suppressing (possible) infrared
singularities of photon legs. The two-point photon vertex has the following
structure:

        \be
        \Pi_{\mu\nu}=(p^2 g_{\mu\nu}-p_\mu p_\nu) A
        +\i \ve_{\mu\nu\lambda}p^\lambda B~~~,
        \ee

\noi
with $A$ and $B$ having finite limits at $p \to 0$. Therefore, dressed photon
lines have the same linear infrared singularity as the bare one, and
any insertion of fermion loops is infrared safe.

Nonclosed internal fermion lines end with external fermion legs.
These (dressed) internal fermion lines may be tied together by photon lines
through (dressed) vertices.  Let us choose  the momenta of internal photon
lines as the integration momenta in the diagram loops. Suppose the loop has
no vertices with external photon legs. In this case, since the photon
propagator brings $1/k$ singularity, infrared divergency in the loop integral
may arise when the loop contains two other lines with $1/k$ singularities.
The additional $1/k$ singularities may come only from the denominators of
neighboring fermion propagators when other terms in these denominators are
cancelled due to the mass shell condition. Therefore, potentially
logarithmically divergent diagrams have the following structure:

\begin{equation}
\int \!dk
\begin{picture}(100.00,40.00)
\unitlength=1.00mm
\linethickness{0.4pt}
\put(-30,-18){
\put(34.00,25.){\line(1,0){30.}}
\put(39.00,25.){\vector(1,0){1.}}
\put(59.00,25.){\vector(1,0){1.}}
\put(34.00,15.){\line(1,0){30.}}
\put(39.00,15.){\vector(1,0){1.}}
\put(59.00,15.){\vector(1,0){1.}}
\put(70.00,20.00){\oval(15.,15.)}
\put(74.00,27.){\line(1,1){7.}}
\put(78.00,20.){\line(1,0){10.}}
\put(74.00,13.){\line(1,-1){7.}}
\put(44,15.75){\oval(1.5,1.5)[r]}
\put(44,17.25){\oval(1.5,1.5)[l]}
\put(44,18.75){\oval(1.5,1.5)[r]}
\put(44,20.25){\oval(1.5,1.5)[l]}
\put(44,21.75){\oval(1.5,1.5)[r]}
\put(44,23.25){\oval(1.5,1.5)[l]}
\put(44,24.75){\oval(1.5,1.5)[br]}
\put(45.,25.){\circle*{1.}}
\put(44.,15.){\circle*{1.}}
\put(37.00,28.){\makebox(0,0)[cc]{{\small $p_1,\a_1$}}}
\put(37.00,12.){\makebox(0,0)[cc]{{\small $p_2,\a_2$}}}
\put(70.00,20.){\makebox(0,0)[cc]{$T$}}
\put(48.00,20.){\makebox(0,0)[cc]{{\small $k$}}}
\put(55.00,28.){\makebox(0,0)[cc]{{\small $p_1+k$}}}
\put(55.00,12.){\makebox(0,0)[cc]{{\small $p_2-k$}}}
}
\end{picture}~~~~~~~~~~~~~~~~~~~~~~~~~~.
\label{fig3}
\end{equation}

\vspace{1cm}
\noi
The infrared-dangerous part of \Eq{fig3} is equal to

        \be
        \int \!dk
        \frac{(\gamma^\mu(\hp_1+M))_{\a_1\beta_1}
        (\gamma^\nu(\hp_2+M))_{\a_2\beta_2}
        \ve_{\mu\nu\l}k^\l}
        {((p_1+k)^2-M^2)((p_2-k)^2-M^2)k^2}
        T_{\beta_1\beta_2}~~~.
        \label{diverg1}
        \ee

\noi
In the vicinity of the fermion mass shell, i.e. when
~$\hp - M= {\de}/{2M}$, ~\Eq{diverg1} is proportional to $\de \ln \de$
and vanishes on the fermion mass shell, $\de\to0$.  Naively, the dressing
of the fermion lines in \Eq{fig3} with external soft photons would
worsen the diagram. However, additional fermion propagators are
infrared-regular.  For instance, let a photon with momentum $p$ be
emitted from the upper fermion line, \Eq{fig3}. The denominator of the
additional fermion propagator is equal to

        \be
        (p_1+k-p)^2-M^2=k^2-2pk+p^2+\de -2p_1p+2kp_1
        \ee

\noi
and it is nonzero even at $p^2=0$.

        Thus we have demonstrated that the anomaly is absent from the
topologically massive QED$_{2+1}$ and calculations may be performed
in any relativistic gauge $\a$ provided that proper regularization is done.
Detailed analysis of the problem of the fermion mass shell gauge
invariance will be presented elsewhere \cite{TZ}.

\section*{Acknowledgement}
We are grateful to A.~E.~Shabad for fruitful discussions.
This work was supported in part by grants RFBR 96-02-17314 and INTAS-RFBR
95-829 (I.~T.) and  RFBR 96-02-16287 (V.~Z.).

\end{document}